Distribution of *s*-, *r*-, and *p*-process nuclides in the early Solar System inferred from Sr isotope anomalies in meteorites

Jonas M. Schneider[1,2]*, Christoph Burkhardt[1,2] and Thorsten Kleine[1,2]

[1]Max Planck Institute for Solar System Research, Justus-von-Liebig-Weg 3, D-37077 Göttingen, Germany

[2]Institute for Planetology, University of Münster, Wilhelm-Klemm-Straße 10, D-48149 Münster, Germany

*corresponding author:

e-mail: *schneiderj@mps.mpg.de

phone: +49 551 384979-308

Abstract: 240

Main text: 4035

4 Figures

1 Table

1 Apppendix (2 Tables)

Abstract

Nucleosynthetic isotope anomalies in meteorites allow distinguishing between the non-carbonaceous (NC) and carbonaceous (CC) meteorite reservoirs and show that correlated isotope anomalies exist in both reservoirs. It is debated, however, whether these anomalies reflect thermal processing of presolar dust in the disk or are primordial heterogeneities inherited from the Solar System's parental molecular cloud. Here, using new high-precision $^{84}Sr$ isotope data, we show that NC meteorites, Mars, and the Earth and Moon are characterized by the same $^{84}Sr$ isotopic composition. This $^{84}Sr$ homogeneity of the inner Solar System contrasts with the well-resolved and correlated isotope anomalies among NC meteorites observed for other elements, and most likely reflects correlated *s*- and (*r*-, *p*-)-process heterogeneities leading to $^{84}Sr$ excesses and deficits of similar magnitude which cancel each other. For the same reason there is no clearly resolved $^{84}Sr$ difference between NC and CC meteorites, because in some carbonaceous chondrites the characteristic $^{84}Sr$ excess of the CC reservoir is counterbalanced by an $^{84}Sr$ deficit resulting from *s*-process variations. Nevertheless, most carbonaceous chondrites exhibit $^{84}Sr$ excesses, which reflect admixture of refractory inclusions and more pronounced *s*-process heterogeneities in these samples. Together, the correlated variation of *s*- and (*r*-, *p*-)-process nuclides revealed by the $^{84}Sr$ data of this study refute an origin of these isotope anomalies solely by processing of presolar dust grains, but points to primordial mixing of isotopically distinct dust reservoirs as the dominant process producing the isotopic heterogeneity of the Solar System.

## 1. Introduction

Nucleosynthetic isotope anomalies reflect the heterogeneous distribution of isotopically anomalous presolar material and show that the solar accretion disk can be subdivided into the non-carbonaceous (NC) and carbonaceous (CC) meteorite reservoirs, which represent two spatially separated, but coexisting regions of the disk (Budde et al., 2016; Kruijer et al., 2017; Warren, 2011). Within this framework, the NC reservoir is typically associated with the inner disk, whereas the CC reservoir is thought to represent more distal disk regions and is possibly located beyond the orbit of Jupiter (Budde et al., 2016; Kruijer et al., 2017; Warren, 2011; Morbidelli et al., 2022). The origin of the NC-CC dichotomy, and more generally the processes that led to isotopic heterogeneity among meteorites and planets, are debated and processes associated with material infall from an isotopically heterogeneous molecular cloud core (Dauphas et al., 2002; Nanne et al., 2019; Burkhardt et al., 2019) or processing of presolar dust grains in the disk (e.g., Trinquier et al., 2009; Regelous et al., 2008, Paton et al., 2013) have been proposed. Furthermore, within each reservoir, isotope anomalies are often correlated, which has been interpreted to reflect mixing of isotopically distinct reservoirs in the disk (Alexander, C.M.O'D., 2019; Spitzer et al., 2020; Burkhardt et al., 2021; Hellmann et al., 2023).

The NC-CC isotopic dichotomy is evident for many of the Fe-peak elements (e.g., Ti, Cr, Ni), and also for several of the heavier elements such as Zr, Mo, and Ru (for recent reviews see Kleine et al., 2020; Kruijer et al., 2020; Bermingham et al., 2020). The isotope heterogeneities for these heavier elements are governed by variations in the relative abundances of nuclides produced in the proton-capture (*p*-process) as well as slow (*s*-process) and rapid (*r*-process) neutron-capture processes of stellar nucleosynthesis. Of these elements, Mo is uniquely useful because it can distinguish between *s*-process and *r*-process isotope variations. It shows that whereas *s*-process Mo isotope variations exist within both reservoirs, the CC reservoir is characterized by an *r*-process excess over the NC reservoir (e.g., Budde et al., 2016: Poole et al., 2017). This *r*-process excess has been interpreted to reflect a larger fraction of material in the CC reservoir with an isotopic composition as measured in Ca,Al-rich inclusions (CAIs), most of which display a more pronounced *r*-process Mo excess than the CC reservoir (Burkhardt et al., 2019; Nanne et al., 2019; Brennecka et al., 2020). For other elements such as Zr and Ru, the distinction between *s*-process and *r*-process variations cannot easily be made, and so for these elements the characteristic *r*-process excess of the CC reservoir is partially masked by an on average larger *s*-process deficit in the CC over the NC reservoir (e.g., Akram et al., 2015; Fischer-Gödde and Kleine, 2017; Render et al., 2022).

A key element for further study of the nucleosynthetic isotope heterogeneity and the distribution of *p*-, *s*-, and *r*-process nuclides in the early Solar System is strontium (Sr). Strontium has four stable isotopes ($^{84}$Sr, $^{86}$Sr, $^{87}$Sr, $^{88}$Sr), of which $^{84}$Sr is produced solely in the *p*-process, $^{86}$Sr and $^{87}$Sr are produced in the *s*-process, and $^{88}$Sr is produced by both, the *s*-process and the *r*-process. Furthermore, $^{87}$Sr is the decay-product of $^{87}$Rb, which over the life-time of the solar system produced radiogenic $^{87}$Sr variations that are orders of magnitude larger than any potential nucleosynthetic $^{87}$Sr heterogeneity (e.g., Papanastassiou and Wasserburg, 1978). Variations in $^{87}$Sr/$^{86}$Sr can, therefore, not be used to assess potential nucleosynthetic Sr isotope heterogeneities. As such, nucleosynthetic Sr isotope anomalies are commonly expressed as variations in the $^{84}$Sr/$^{86}$Sr ratio after mass-bias correction by internal normalization to $^{88}$Sr/$^{86}$Sr, and so variations in $^{84}$Sr can reflect heterogeneity in *p*-, *s*- and *r*-process Sr (Papanastassiou and Wasserburg, 1978). This makes nucleosynthetic Sr isotope variations a promising tool to investigate the distribution of *p*-, *r*-, and *s*-process nuclides in the early Solar System.

Nucleosynthetic Sr isotope heterogeneity has been shown for acid leachates of primitive chondrites (Qin et al. 2011; Paton et al. 2013, Burkhardt et al., 2019), CAIs (Moynier et al., 2012; Hans et al., 2013), and several bulk meteorites, where Sr anomalies have predominantly been found for CC meteorites but also some NC meteorites (Fukai & Yokoyama, 2019; Hans et al., 2013; Henshall et al., 2018; Moynier et al., 2012; Paton et al., 2013; Yokoyama et al., 2015). However, overall Sr isotopes seem to show less systematic behavior than observed for the nucleosynthetic isotope anomalies of many other elements. For instance, it is unclear as to whether there is a systematic Sr isotope offset between NC and CC meteorites and whether there is Sr isotope heterogeneity in the NC reservoir. We present new high-precision $^{84}$Sr data for a comprehensive suite of meteorites, including several samples for most of the major NC and CC meteorite groups. The new data are used to unravel the nature and extent of Sr isotope variations among meteorites, to assess the distribution of *p*-, *s*-, and *r*-process nuclides among meteorites and the terrestrial planets, and to constrain the processes responsible for generating the nucleosynthetic isotope heterogeneity in the early Solar System.

## 2. Samples and analytical methods

Thirty-nine samples were selected for this study, including five terrestrial rocks, five lunar samples, and 29 meteorites from various chondrite and achondrite groups (Table 1). To avoid alteration induced by terrestrial weathering (Fukai and Yokoyama, 2019), we preferentially selected meteorite falls (n=19) and finds from Antarctica (n=9) and avoided finds from hot deserts (except NWA 4590).

Carbonaceous chondrites and the terrestrial sample JB2#2 were digested in Parr Bombs for 96h at 190°C using HF-$HNO_3$ (1:1) to facilitate complete digestion of refractory phases. All other samples were digested in Teflon beakers on a hotplate using a HF-$HNO_3$ (3:1) at >140°C for 72h. After digestion, all samples were repeatedly re-dissolved and dried in *aqua regia* to dissolve residual fluorides. Strontium was purified by ion exchange chromatography following established procedures (e.g., Brennecka et al. 2013), with an additional purification step using Eichrom Sr-spec resin.

The Sr isotope measurements were performed using a ThermoScientific Triton *Plus* thermal ionization mass spectrometer (TIMS) at the University of Münster. The $^{84}$Sr/$^{86}$Sr ratios were measured using a two-line dynamic acquisition, where $^{84}$Sr/$^{86}$Sr measured in the first line is corrected for instrumental mass fractionation using $^{86}$Sr/$^{88}$Sr measured in the second line. The $^{84}$Sr/$^{86}$Sr and $^{86}$Sr/$^{88}$Sr ratios measured in this manner use the same Faraday cups, and so any bias induced by different cup efficiencies or amplifier gains cancel out (e.g., Hans et al., 2013; Yobregat et al., 2017). The dynamic $^{84}$Sr/$^{86}$Sr ratios show excellent agreement for all four sessions with a mean $^{84}$Sr/$^{86}$Sr = 0.0564907±21 (2σ) for the NIST SRM 987 standard, whereas the static $^{84}$Sr/$^{86}$Sr measurements exhibit significant drift, consistent with observations of prior studies (Hans et al., 2013; Henshall et al., 2018).

## 3. Nucleosynthetic Sr isotope variations among meteorites

The Sr isotopic data are reported as $\mu^{84}$Sr values, which represents the parts per $10^6$deviation of the mass-bias corrected $^{84}$Sr/$^{86}$Sr in a sample relative to the same ratio in the NIST SRM 987 standard ($\mu^{84}Sr = [(^{84}Sr/^{86}Sr)_{sample}/(^{84}Sr/^{86}Sr)_{SRM987} - 1] \times 10^6$). The $\mu^{84}$Sr values of the terrestrial samples are indistinguishable from one another (average $\mu^{84}$Sr = 2±12, 2 s.d.) and from the NIST SRM 987 standard (Table 1, Fig. 1). This demonstrates the accuracy of our analytical methods and implies that the SRM 987 standard provides a good proxy for the $\mu^{84}$Sr of bulk Earth. This is consistent with prior studies that also employed a dynamic measurement routine (Hans et al. 2013, Yobregat et al. 2017, Henshall et al. 2018), but differs from results of some other studies, which based on static measurements found negative $\mu^{84}$Sr of approximately –20 ppm for terrestrial samples (e.g., Moynier et al. 2012; Paton et al., 2013). This difference can likely be attributed to the larger uncertainty inherent in the static Sr isotope measurements and highlights the importance of using dynamic measurements to reliably resolve the anticipated small $\mu^{84}$Sr variations among meteorites.

The new Sr isotope data show that different samples from a given meteorite group have indistinguishable $\mu^{84}$Sr values, which makes it possible to calculate precise group means (Table 1; Fig. 1). The five lunar samples display a mean $\mu^{84}$Sr of –1±10, indicating that the Earth and Moon share the same $\mu^{84}$Sr. The four martian meteorites define a mean $\mu^{84}$Sr of 12±12, which again is indistinguishable from the terrestrial and lunar values. Finally, for the four major NC meteorite groups investigated in this study (enstatite and ordinary chondrites, angrites, eucrites), all samples are indistinguishable from one another and provide precise mean $\mu^{84}$Sr values for each group as well a mean $\mu^{84}$Sr = 6±5 (95% conf.) for the NC reservoir. This mean value overlaps with the composition of Earth, Moon, and Mars, and including these compositions results in a precise mean inner Solar System $\mu^{84}$Sr value of 5±4 (95% conf.). Thus, the inner Solar System appears to be homogeneous with respect to $\mu^{84}$Sr. This contrasts with results of some prior studies, which reported $\mu^{84}$Sr variations among NC meteorites of up to ~30 ppm (Moynier et al., 2012; Paton et al., 2013). For example, Moynier et al. (2012) reported a $\mu^{84}$Sr of –36±6 for enstatite chondrites based on static measurements of a single sample, and found a range of $\mu^{84}$Sr for ordinary chondrites from –43±10 to –4±5. As for the terrestrial samples, the disparity between some of the previous data and the data of this study for NC materials most likely reflects the higher precision of the dynamic measurement routine combined with the more comprehensive sample set used in the present study.

Most carbonaceous chondrites of this study exhibit more anomalous $\mu^{84}$Sr values but partly overlap with the NC composition (Fig. 1). Whereas CV, CM, and CR have $^{84}$Sr excesses of between ~64 and ~78 ppm, the CI chondrite Orgueil and the two ungrouped chondrites Tagish Lake (TL) and Tarda exhibit lower $\mu^{84}$Sr that partly overlap with those of NC meteorites (Table 1, Fig. 1). This agrees with results of prior studies, which also showed that carbonaceous chondrites display variable $^{84}$Sr excesses (e.g., Yokoyama et al., 2015; Fukai and Yokoyama, 2019). However, the $\mu^{84}$Sr value for CR chondrites of this study is more elevated (mean $\mu^{84}$Sr = 73±33) than $\mu^{84}$Sr values of ~0 reported for two CR chondrite desert finds in earlier studies (Moynier et al., 2012; Fukai and Yokoyama, 2019). This difference almost certainly reflects terrestrial alteration in these desert finds (Fukai and Yokoyama, 2019), which is absent in the CR chondrites of this study, all of which are finds from Antarctica. Finally, for the CI chondrite Orgueil, $\mu^{84}$Sr = 20±22 determined here agrees with $\mu^{84}$Sr = –1±25 for Orgueil and 18±25 for the CI chondrite Yamato 980115 (Fukai and Yokoyama, 2019), but is slightly lower than, albeit not resolved from $\mu^{84}$Sr = 42±17 reported for Orgueil (Moynier et al., 2012) and $\mu^{84}$Sr = 37±10 reported for the CI chondrite Ivuna (Paton et al., 2013). These slightly higher $\mu^{84}$Sr values probably reflect incomplete digestion of acid-resistant presolar grains such as SiC (Fukai & Yokoyama 2019).

## 4. Origin of $^{84}$Sr variations in the Solar System

The new data of this study reveal that the $^{84}$Sr systematics of meteorites differ in two important ways from other elements. First, there are no resolved $^{84}$Sr variations within the NC reservoir. This differs from the isotope anomalies for many other elements, which exhibit well-resolved and correlated isotope variations among NC meteorites, Mars, and the Earth and Moon. Given this ubiquitous isotope heterogeneity among NC meteorites, which is present for all elements showing nucleosynthetic isotope anomalies in carbonaceous chondrites or their components, it appears highly unlikely that $^{84}$Sr is indeed homogeneous in the NC reservoir. Second, the $\mu^{84}$Sr values of NC and CC meteorites are not separated by a gap, but partly overlap. This contrast with several other elements, where NC and CC meteorites define two distinct compositional clusters. As we will show below, this contrasting behavior of Sr likely results from the superimposed effects of (*p*-, *r*-) and *s*-process heterogeneities, which reduce the overall $^{84}$Sr variations within and between the NC and CC reservoirs.

### *4.1. $^{84}$Sr homogeneity of the inner Solar System*

The nucleosynthetic isotope heterogeneity among meteorites is governed by isotope anomalies in the Fe-peak elements (Ca, Ti, Cr, Ni, Zn), which are mainly produced by nuclear statistical equilibrium (NSE) and explosive nucleosynthesis in (pre-)supernovae massive stars, and anomalies in heavier elements such as Zr, Mo, and Ru associated with variations in the abundance of *p*-, *r*-, and *s*-process nuclides (e.g. Woosley and Heger, 2002, Pignatari et al., 2016). Whereas the CC reservoir is characterized by excesses in nuclides produced in neutron-rich stellar environments (e.g., $^{50}$Ti, $^{54}$Cr, *r*-process Mo), the NC reservoir exhibits deficits in these nuclides, relative to the terrestrial standard (e.g., Kleine et al., 2020). There are additional *s*-process variations in both reservoirs, which in the NC reservoir are correlated with the isotope anomalies in the Fe-peak elements and the variation in the relative abundance of *r*-process Mo (e.g., Spitzer et al., 2020). Thus, $\mu^{84}$Sr variations among and between NC and CC meteorites might be expected due to (*i*) *s*-process variations in both reservoirs, (*ii*) the *r*-process NC-CC difference, and (*iii*) *r*-process variations in the NC reservoir. Understanding the extent of $^{84}$Sr variations among meteorites, therefore, requires disentangling the individual contributions of the specific *s*-, *r*-, and *p*-process variations to the $\mu^{84}$Sr value of any given sample.

To quantitatively assess the expected $\mu^{84}$Sr variations, we will start with the variations expected from the heterogeneous distribution of nuclides produced in neutron-rich stellar environments. For Sr this includes *r*- and *p*-process variations, which result in the same $^{84}$Sr signatures; for instance, the $^{84}$Sr excesses in CAIs have been interpreted to reflect either an *r*-process (Hans et al., 2013) or *p*-process excess (Charlier et al., 2021). For the model of this study, the distinction between *r*- and *p*-process variations is not important, which we will therefore treat together. We begin with the observation that the isotopic composition of the CC reservoir is always intermediate between those of the NC reservoir and CAIs (Nanne et al., 2019; Burkhardt et al., 2019). This observation has led to a model which attributes the ultimate origin of the NC-CC isotopic difference to time-varied infall from an isotopically heterogeneous molecular cloud, followed by mixing within the disk (Nanne et al., 2019; Burkhardt et al., 2019; Jacquet et al., 2019). In this model, CAIs represent the isotopic composition of the early infall, while NC meteorites record the isotopic composition of the late infall. The isotopic composition of the CC reservoir is then attributed to mixing of NC material with material having a CAI-like isotopic but broadly chondritic bulk chemical composition (termed IC reservoir for Inclusion-like Chondritic reservoir; Burkhardt et al., 2019). The nucleosynthetic variations within the NC reservoir are on the same trajectory as the NC-CC-IC offset, and so both the NC-CC isotopic difference as well as the isotopic variations within the NC reservoir can be understood as the result of mixing between the same two reservoirs, namely the IC reservoir and a reservoir having the starting composition of the NC reservoir (Fig. 2a). The fraction of IC material in the NC reservoir is then given by mass balance:

$$f_{IC,NC} = \frac{\mu_{NC,high} - \mu_{NC,low}}{\mu_{IC} - \mu_{NC,low}} \quad (1)$$

where $\mu_{IC}$ is given by CAIs, and $\mu_{NC,high}$ and $\mu_{NC,low}$ represent the isotopic compositions of the two extremes of the NC isotope trend. The value of $f_{IC,NC}$ can be calculated for those elements whose isotopic composition in CAIs is well known and which exhibit sufficient isotope variation among NC meteorites. This includes $^{48}$Ca, $^{50}$Ti, $^{54}$Cr, and Mo. For Mo, this requires isolating the *s*- and *r*-process variations from each other. For this purpose, Budde et al. (2019) introduced the $\Delta^{95}$Mo notation, which provides the deviation of a sample's Mo isotopic composition from a theoretical *s*-process mixing line and, as such, is a measure of *r*-process Mo variations. All four elements provide consistent $f_{IC,NC}$ values with a weighted mean of 0.16±0.02 (Fig. 3, Appendix Table A1&A2).

Earth typically plots at one end of the NC isotope trend, defining the value of $\mu_{NC,high}$ in equation (1). As μ-values are calculated relative to the composition of the terrestrial standard and because the isotopic composition of Earth is typically close to that of the standard, we will for simplicity assume $\mu_{NC,high} \approx 0$. Rearranging equation (1) we have:

$$\mu_{NC,low} = \frac{-f_{IC,NC} \times \mu_{IC}}{1 - f_{IC,NC}} \quad (2)$$

Using the mean $\mu^{84}$Sr = 131±11 (95% conf.) of CAIs (Brennecka et al. 2013, Hans et al. 2013, Paton et al., 2013, Myoyo et al. 2018, Brennecka et al. 2020, Charlier et al. 2021) for $\mu_{IC}$ and $f_{IC,NC}$ = 0.16±0.02 yields an expected $\mu^{84}Sr_{NC,low}$ value of –24±6. In other words, *without s*-process variations the NC reservoir would be expected to have a range of $\mu^{84}$Sr values from –24±6 for the most (*r*-, *p*-)-process depleted bodies (e.g., ureilites) to ~0 for enstatite chondrites and Earth (Fig. 4, Table A2).

However, in addition to (*r*-, *p*-)-process variations, NC meteorites also display correlated *s*-process variations, which range from only small *s*-process deficits in enstatite chondrites to larger deficits in for instance IIAB iron meteorites (e.g., Budde et al., 2019; Spitzer et al., 2020). To quantify the effect of these *s*-process variations on $\mu^{84}$Sr, we will take Mo as a proxy. The $\mu^{94}$Mo variations among NC meteorites are dominated by *s*-process variations with $\mu^{94}$Mo values ranging from ~0 for IAB iron meteorites and winonaites to ~120 for IIAB iron meteorites (Budde et al. 2019, Spitzer et al., 2020). On the basis of Mo and Sr isotopic data for presolar SiC grains (Nicolussi et al., 1998), acid leachates of primitive chondrites (Burkhardt et al., 2019), and *s*-process theory (Arlandini et al., 1999, Bisterzo et al., 2014, Dauphas et al., 2014), we estimate that *s*-process heterogeneity results in 0.31±0.07 times smaller $\mu^{84}$Sr than $\mu^{94}$Mo variations (Burkhardt et al. 2019). Consequently, an *s*-process deficit of $\mu^{94}$Mo = 120 corresponds to an expected *s*-process $\mu^{84}$Sr anomaly of 36±10. This expected *s*-process $\mu^{84}$Sr anomaly is of similar magnitude but opposite sign than the corresponding expected (*r*-, *p*-)-process $\mu^{84}$Sr anomaly of –24±6 from above. Due to the correlated nature of the *s*- and *r*-process variations among NC meteorites (Fig. 2b), samples with the largest (*r*-, *p*-)-process effects on $\mu^{84}$Sr also have the largest *s*-process anomaly, while samples with the smallest (*r*-, *p*-)-process anomaly also have the smallest *s*-process anomaly. Thus, for all NC samples the resulting positive (*s*-process) and negative (*r*-, *p*-process) $\mu^{84}$Sr anomalies cancel out, which leads to the observed $\mu^{84}$Sr values close to 0 throughout the NC reservoir (Fig. 4).

### *4.2. $^{84}$Sr offset between NC and CC meteorites*

The model from above can also be used to assess the expected $\mu^{84}$Sr difference between NC and CC meteorites. For a refractory element like Sr, the Sr isotopic composition of carbonaceous chondrites can be strongly affected by the admixture of CAIs, and so for assessing the characteristic offset between the NC and CC reservoirs it is useful to use carbonaceous chondrites which contain only little CAIs, such as CI chondrites. For the CC reservoir, equation (1) takes the form:

$$f_{IC,CC} = \frac{\mu_{CI} - \mu_{NC,low}}{\mu_{IC} - \mu_{NC,low}} \tag{3}$$

Similar to the calculation for the NC reservoir, values for $f_{IC,CC}$ can be calculated using data for $^{48}$Ca, $^{50}$Ti, $^{54}$Cr, and $\Delta^{95}$Mo, and using the composition of CAIs to represent $\mu_{IC}$ and the composition of ureilites to represent $\mu_{NC,low}$ (Table A1). We find a mean $f_{IC,CC}$ = 0.36±0.03 (Fig. 3), which can be used to calculate the expected $\mu^{84}$Sr for CI chondrites:

$$\mu_{CI} = f_{IC,CC} \times \mu_{IC} + (1 - f_{IC,CC}) \times \mu_{NC,low} \tag{4}$$

Using the mean $\mu^{84}$Sr = 131±11 of CAIs for $\mu_{IC}$, $f_{IC,CC}$ = 0.36±0.06, and $\mu^{84}$Sr = –24±6 for $\mu_{NC,low}$ provides an expected $\mu^{84}$Sr value for CI chondrites of 32±16. This value is calculated solely based on the expected (*r*-, *p*-)-process $\mu^{84}$Sr variations, but there are additional *s*-process variations affecting the $^{84}$Sr isotopic composition of carbonaceous chondrites. To assess the magnitude of these *s*-process variations, we again take $^{94}$Mo as a proxy. The $\mu^{94}$Mo values of IIAB irons (which have the largest *s*-process deficit and hence define $\mu_{NC,low}$ for Mo) and CI chondrites differ by ~40 ppm ($\mu^{94}$Mo ≈120 for IIAB irons and ≈80 for CI chondrites), which predominantly reflect *s*-process variations. Using the 0.31 ratio between $\mu^{84}$Sr and $\mu^{94}$Mo *s*-process variations implies that the expected $\mu^{84}$Sr for CI chondrites needs to be lowered by ~12 ppm, resulting in an overall expected $\mu^{84}$Sr = 20±17. This value overlaps with the measured $\mu^{84}$Sr = 20±22 for Orgueil from this study, indicating that the lack of a clearly resolved $\mu^{84}$Sr difference between NC and CC meteorites reflects the counteracting effects of *s*- and *r*-process variations.

### *4.3. Origin of $^{84}$Sr variations among carbonaceous chondrites*

Prior studies have shown that the variable abundances of CAIs exert a strong control on the $^{84}$Sr isotopic composition of bulk carbonaceous chondrites (Fukai & Yokoyama 2018, Burkhardt et al. 2019), which is consistent with the elevated $\mu^{84}$Sr values of the relatively CAI-rich CV chondrites reported in this and prior studies. However, this study reports the first precise $^{84}$Sr data for CR chondrites, which contain only few CAIs but display the same $\mu^{84}$Sr value as the CAI-rich CV chondrites. Consequently, variable abundances of CAIs cannot be the sole cause of $\mu^{84}$Sr variations among the carbonaceous chondrites. CR chondrites are characterized by one of the largest *s*-process deficits with a $\mu^{94}$Mo value of ~280 (Budde et al., 2018). For comparison, CI chondrites display $\mu^{94}$Mo ≈ 80, and so there is a ~200 ppm $\mu^{94}$Mo difference between the CR and CI chondrites. This corresponds to an *s*-process-induced $\mu^{84}$Sr difference between these two carbonaceous chondrites of ~60 ppm (for *s*-process variations $\mu^{84}$Sr ≈ 0.31 × $\mu^{94}$Mo; see above) and fully accounts for the more elevated $\mu^{84}$Sr of ~73 for CR chondrites compared to CI chondrites ($\mu^{84}$Sr ≈ 20). Thus, not all the $^{84}$Sr variations among carbonaceous chondrites are attributable to the heterogeneous distribution of CAIs, but may also reflect *s*-process variations.

## 5. Implications for the origin of nucleosynthetic variability in the disk

It has been suggested that the nucleosynthetic isotopic heterogeneity of the solar protoplanetary disk reflects thermal processing of presolar dust grains (e.g., Trinquier et al. 2009, Paton et al. 2013). For instance, on the basis of apparent negative $\mu^{84}$Sr values for some NC meteorites and Earth (relative to the terrestrial standard), Paton et al. (2013) argued that inner Solar System bodies are enriched in *s*-process Sr, which is carried by presolar SiC grains, and where the SiC enrichment results from the removal of thermally labile components by thermal processing of dust in the inner disk. However, the new $^{84}$Sr data of this study reveal that neither there are $^{84}$Sr variations among inner Solar System bodies, nor are any of these bodies characterized by negative $\mu^{84}$Sr values (Fig. 1). This strongly argues against thermal processing as a mechanism to produce the nucleosynthetic isotope heterogeneity of the disk, because the resulting enrichment in thermally resistant carriers such as SiC would, as shown by Paton et al. (2013), result in variable $\mu^{84}$Sr among NC meteorites.

In addition, the lack of resolvable $^{84}$Sr variations among NC meteorites indicates correlated heterogeneity of *s*-process- *and* (*r*-, *p*-)-process-derived Sr. Producing such correlated isotope variations by thermal processing of presolar dust grains is highly unlikely, because any given *s*-process excess resulting from thermal processing would have to be balanced by the appropriate excess in (*r*-, *p*-)-process carriers. In other words, the thermal processing would have to fortuitously result in coupled enrichments of distinct presolar carriers such that no resolvable $^{84}$Sr anomaly is produced. While this may happen for individual samples by chance, it seems highly unlikely that this consistently occurred for all NC meteorites and resulted in the observed variable and correlated *s*-process and (*r*-, *p*-)-process isotope variations. This is in line with the observation that among NC meteorites correlated isotope variations exist for elements of different stellar origins (supernovae, AGB stars) and having distinct geochemical (lithophile, siderophile) and cosmochemical (refractory, volatile) properties (Spitzer et al., 2020; Burkhardt et al., 2021). Such multi-elemental correlations of isotope anomalies cannot reflect thermal processing of distinct presolar carriers in the disk, because different elements and carrier phases are expected to have reacted differently to such processing. As such, the results of this study support models in which the nucleosynthetic isotope heterogeneity of the disk is inherited from the Solar System's parental molecular cloud, with additional variations resulting from subsequent mixing within the disk.

## Acknowledgments

We gratefully acknowledge NASA, the Museum National d'histoire Naturelle de Paris, the National History Museum London, and the Senckenbergmuseum Frankfurt for providing samples. US Antarctic meteorite samples are recovered by the Antarctic Search for Meteorites (ANSMET) program which has been funded by NSF and NASA, and characterized and curated by the Department of Mineral Sciences of the Smithsonian Institution and Astromaterials Curation Office at NASA Johnson Space Center. We thank the anonymous referee for a constructive review, which helped to clarify important points and led to improvement of the manuscript. This work was funded by the Deutsche Forschungsgemeinschaft (DFG, German Research Foundation) (Project- ID 263649064-TRR170) and the European Research Council Advanced Grant Holy Earth (grant no. 101019380). This is TRR 170 pub. no. 194.

**Table 1**

Sr isotope data for terrestrial and lunar samples and meteorites.

| Group | Sample | Type | n | $\mu^{84}Sr$ | ± | 2σ |
|---|---|---|---|---|---|---|
| Earth | AGV-2 | Andesite | 14 | 0 | ± | 7 |
| | BCR-2 | Intraplate basalt | 7 | -1 | ± | 12 |
| | BCR-2#2 | Intraplate basalt | 6 | 1 | ± | 20 |
| | BCR-2#3 | Intraplate basalt | 10 | 13 | ± | 14 |
| | BCR-2#4 | Intraplate basalt | 5 | 3 | ± | 28 |
| | BHVO-2 | Ocean island basalt | 8 | 2 | ± | 9 |
| | BHVO-2#4 | Ocean island basalt | 5 | -11 | ± | 22 |
| | JB-2 | Ocean island basalt | 8 | 4 | ± | 21 |
| | JB2#2 | Ocean island basalt | 4 | -3 | ± | 11 |
| | BIR1a | Mid-ocean ridge basalt | 7 | 2 | ± | 14 |
| | *Average* | | | *2* | ± | *4* |
| Moon | 60025.842 (241) | Ferroan anorthosite | 8 | -12 | ± | 14 |
| | 10057 | High-Ti mare basalt | 8 | 5 | ± | 14 |
| | 15495 | Porphyritic basalt | 5 | 5 | ± | 16 |
| | 70017.825 | High-Ti mare basalt | 9 | 3 | ± | 18 |
| | 12002.58 | Porphyritic basalt | 9 | -7 | ± | 16 |
| | *Average* | | | *-1* | ± | *10* |
| Mars | Zagami | Enriched shergottite | 6 | 10 | ± | 23 |
| | MIL03346.76 | Nakhlite | 10 | 17 | ± | 14 |
| | RBT04262.75 | Enriched lherzolitic shergottite | 7 | 2 | ± | 11 |
| | Tissint | Enriched shergottite | 10 | 18 | ± | 6 |
| | *Average* | | | *12* | ± | *12* |
| Eucrites | Bouvante | Monomict eucrite | 6 | 2 | ± | 15 |
| | Juvinas | Monomict eucrite | 5 | 3 | ± | 15 |
| | Millbillilie | Polymict eucrite | 7 | 4 | ± | 22 |
| | Moore County | Cumulate eucrite | 4 | 9 | ± | 34 |
| | Pasamonte | Polymict eucrite | 6 | 5 | ± | 20 |
| | Stannern | Monomict eucrite | 7 | 11 | ± | 11 |
| | Stannern#2 | Monomict eucrite | 7 | 2 | ± | 11 |
| | *Average* | | | *5* | ± | *3* |
| Angrites | D`Orbigny | Volcanic | 6 | 14 | ± | 18 |
| | D'Orbigny#2 | Volcanic | 4 | 13 | ± | 27 |
| | D'Orbigny#3 | Volcanic | 8 | 12 | ± | 14 |
| | NWA 4590 | Sub-volcanic | 4 | 4 | ± | 30 |
| | *Average* | | | *11* | ± | *7* |
| Ordinary Chondrites | Barwell | L5 | 6 | -4 | ± | 23 |
| | Aumale | L6 | 6 | 8 | ± | 16 |
| | Leedey | L6 | 7 | 7 | ± | 12 |
| | *Average* | | | *4* | ± | *16* |
| Enstatite Chondrites | ALHA 81021 | EL6 | 9 | 5 | ± | 13 |
| | LAP10014.22 | EL6 | 4 | -7 | ± | 3 |
| | Khairpur | EL6 | 8 | 10 | ± | 10 |
| | Pillistfer | EL6 | 7 | 4 | ± | 20 |
| | *Average* | | | *3* | ± | *11* |
| Carbonaceous Chondrites | Murchison | CM2 | 7 | 66 | ± | 11 |
| | Murchison#2 | CM2 | 6 | 67 | ± | 10 |
| | Murchison#3 | CM2 | 7 | 63 | ± | 11 |
| | MET 01070 | CM1 | 6 | 60 | ± | 17 |
| | *Average* | | | *64* | ± | *3* |
| | Allende | CV3 | 7 | 91 | ± | 19 |
| | Allende#2 | CV3 | 9 | 65 | ± | 15 |
| | *Average* | | | *78* | ± | *37* |
| | MIL09001 | CR2 | 5 | 72 | ± | 29 |
| | DOM08006 | CR2 | 6 | 70 | ± | 15 |
| | GRO95577 | CR1 | 6 | 99 | ± | 14 |
| | GRA06100.58 | CR2 | 5 | 70 | ± | 46 |
| | GRA06100.58#2 | CR2 | 7 | 53 | ± | 18 |
| | *Average* | | | *73* | ± | *14* |
| | Tarda | C2 ungr. | 7 | 40 | ± | 16 |
| | Tagish Lake | C2 ungr. | 7 | 9 | ± | 12 |
| | Orgueil | CI1 | 5 | 20 | ± | 22 |

**Notes**. $\mu^{84}Sr=(^{84}Sr/^{86}Sr_{sample} / ^{84}Sr/^{86}Sr_{SRM987} - 1) \times 10^6$. *Instrumental mass fractionation was corrected using by internal normalization to* $^{88}Sr/^{86}Sr = 8.3753$ *and using the exponential law. n denotes number of individual measurements. Uncertainties are 95% conf. for n≥4 and 2 s.d. for n<4.*

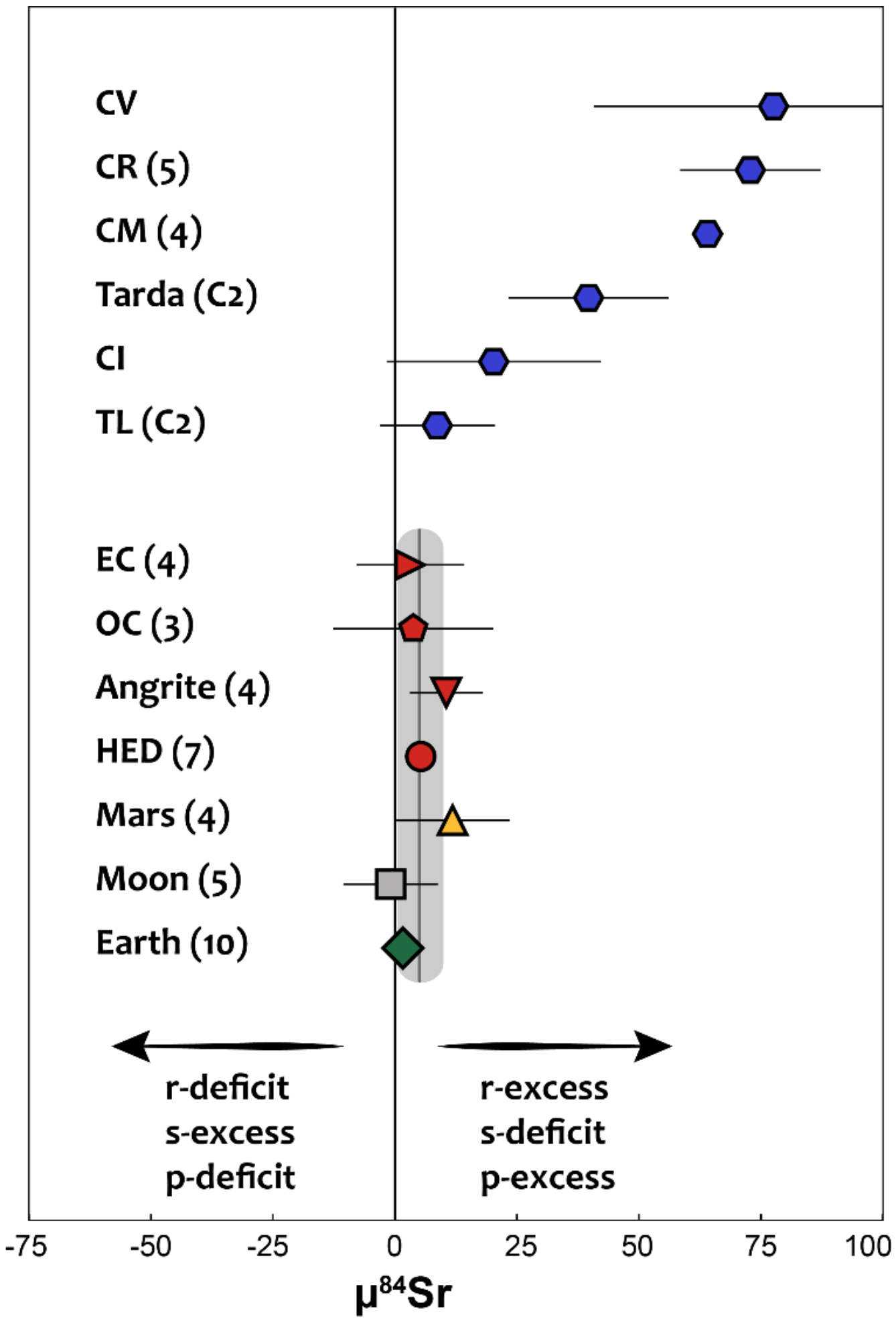

**Figure 1:** Average $\mu^{84}$Sr values of terrestrial, lunar, and meteoritic samples measured in this study. Uncertainties are 95% conf. for n≥4 and 2 s.d. for n<4. Grey box represents inner Solar System average of $\mu^{84}$Sr = 5±4 (95% conf.). Arrows indicate effect of excesses and deficits in *s*-, *r*-, and *p*-process Sr on $\mu^{84}$Sr.

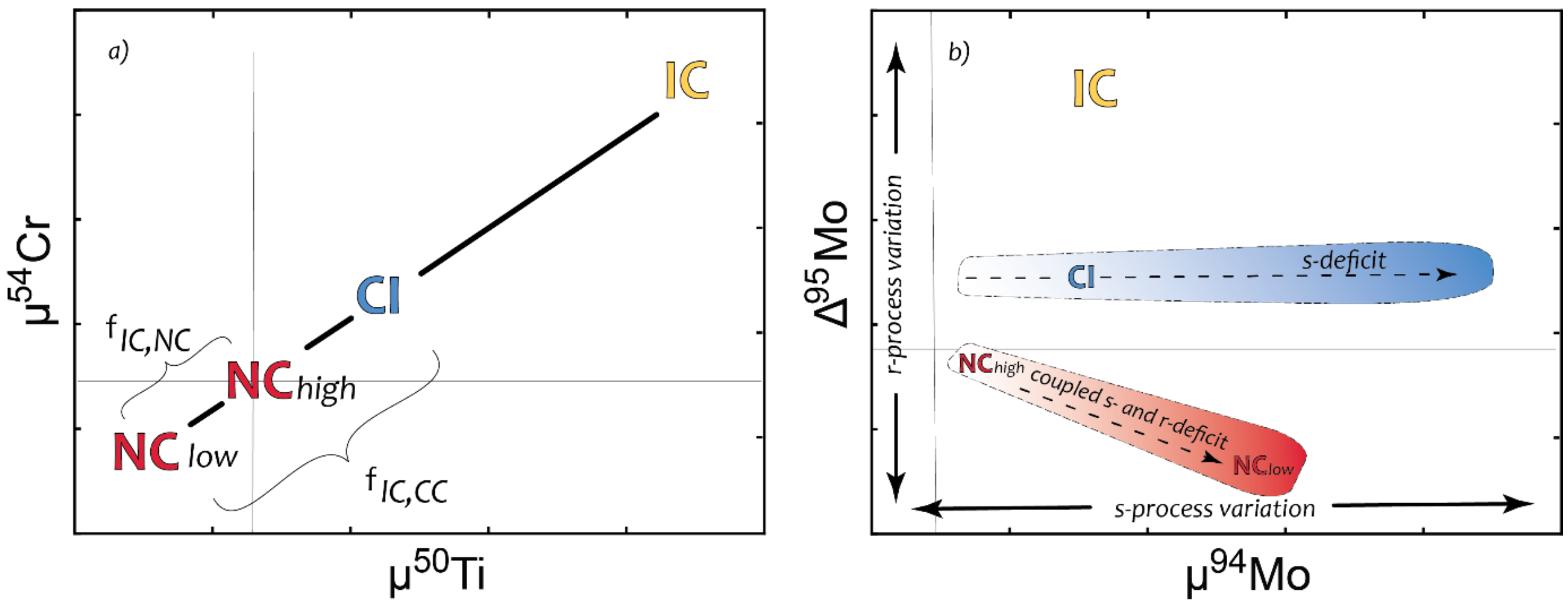

**Figure 2:** Schematic plots illustrating the mixing model of this study. a) Plot of $\mu^{50}Ti$ vs. $\mu^{54}Cr$ showing the location of the three components used in the mixing model: *IC*, $NC_{low}$, $NC_{high}$. Fractions $f_{IC,NC}$ *and* $f_{IC,CC}$ can be calculated by assuming that $NC_{high}$ and *CI* are mixtures of $NC_{low}$ and *IC* (see text for details). b) Plot of $\mu^{94}Mo$ vs. $\Delta^{95}Mo$. For definition of $\Delta^{95}Mo$ see text. Variations in $\Delta^{95}Mo$ reflect *r*-process variations, whereas variations in $\mu^{94}Mo$ mostly reflect *s*-process variation. The position of the mixing endmembers *IC* and $NC_{low}$ are indicated, as well as the resulting compositions of $NC_{high}$ and *CI*. For details about the model see text.

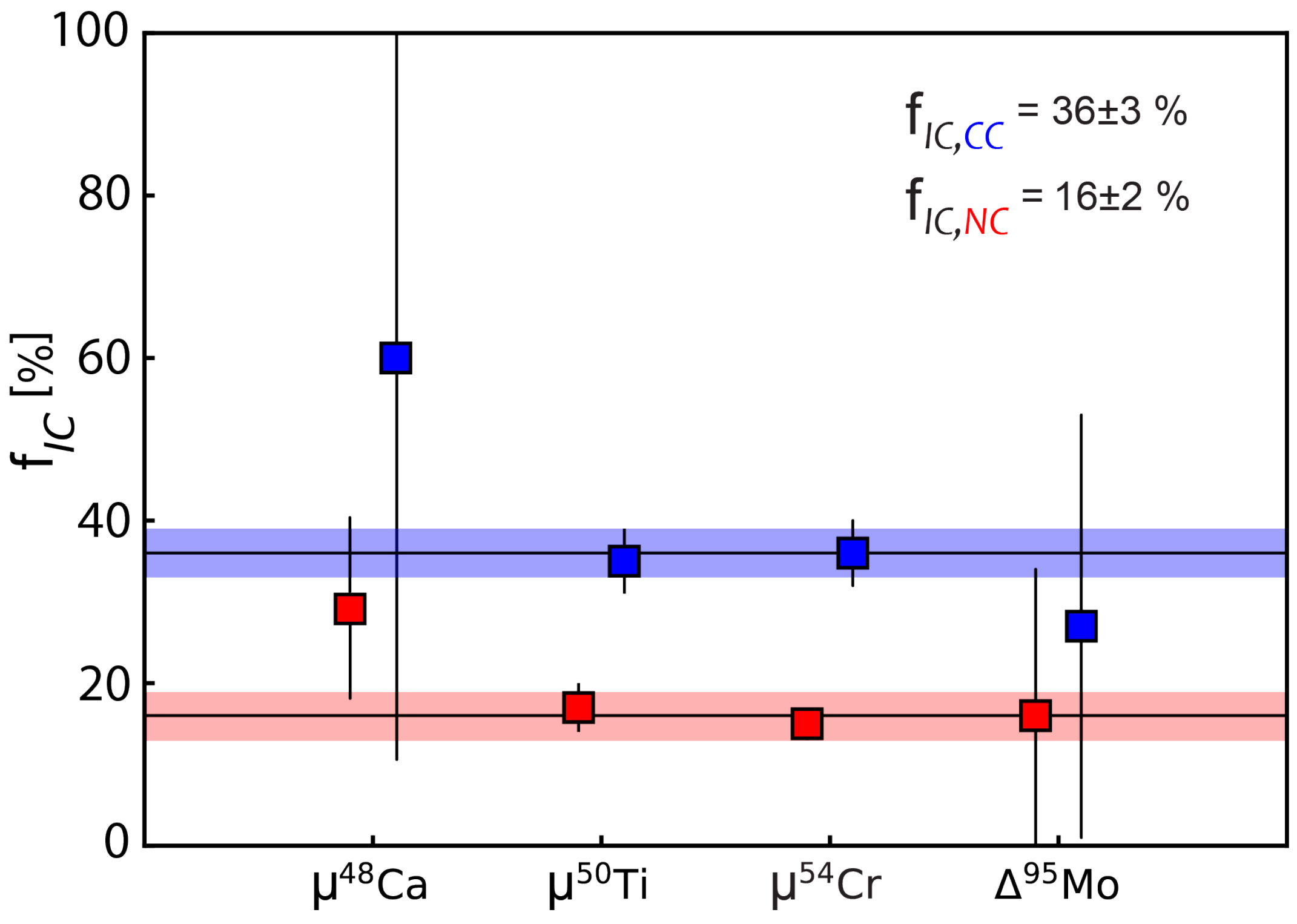

**Figure 3:** Fraction of IC material in the NC and CC reservoirs calculated using equations (1) and (3). Data used in these calculations are summarized in Table A1.

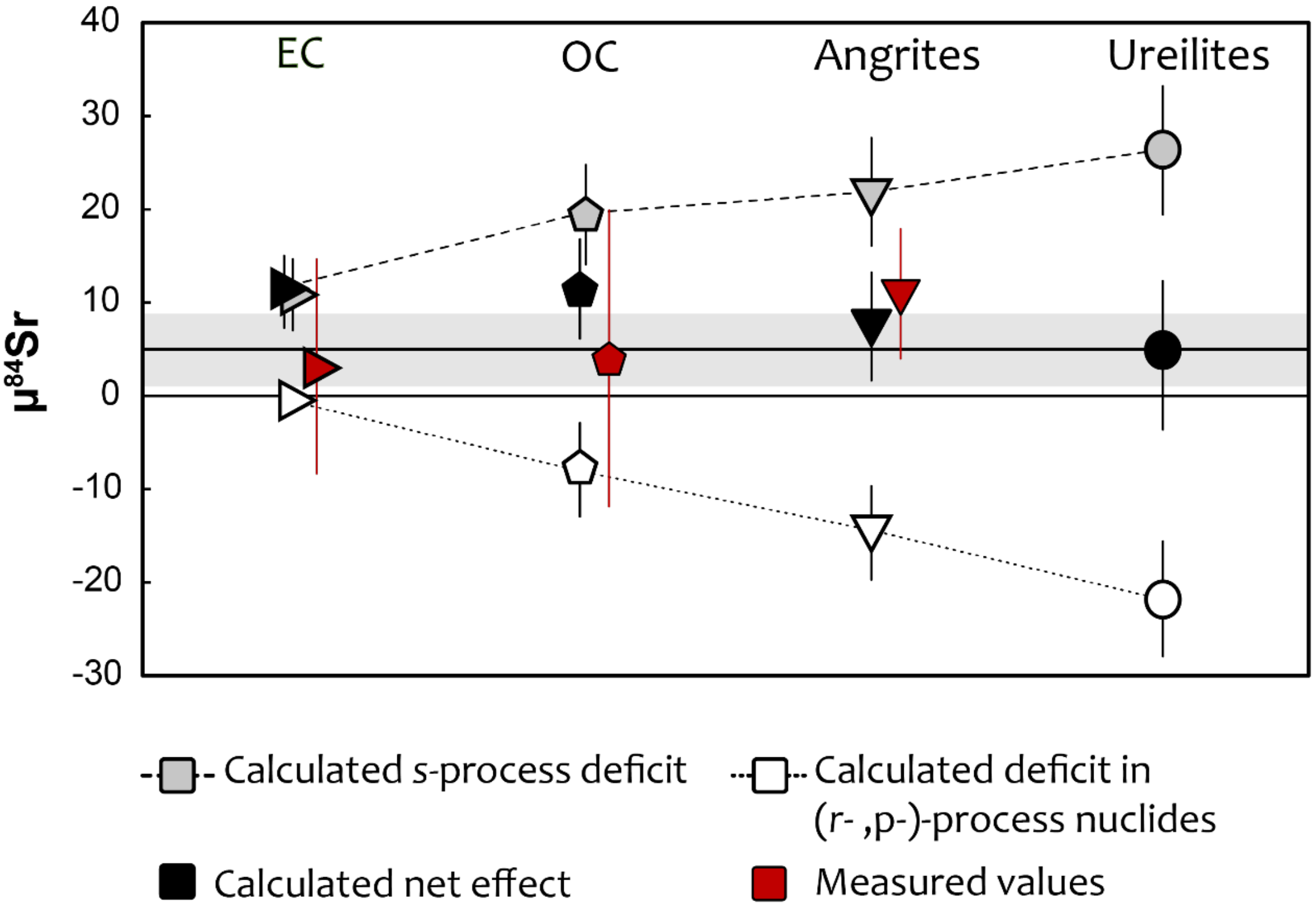

**Figure 4:** Expected $\mu^{84}$Sr anomalies for enstatite chondrites (EC), ordinary chondrites (OC), angrites, and ureilites resulting from only *s*- and only (*r*-, *p-)*-process variations. These variations lead to $\mu^{84}$Sr variations of opposite sign but similar magnitude, and so no resolvable $\mu^{84}$Sr variations remain among NC meteorites. Measured $\mu^{84}$Sr values for EC, OC, and angrites are shown for comparison and are well reproduced by the model. There are no $\mu^{84}$Sr data for ureilites, but they are also predicted to have $\mu^{84}Sr \approx 0$. Grey band represents the inner Solar System average $\mu^{84}Sr = 5 \pm 4$ (95% conf.). Data is summarized in Table A2.

**APPENDIX**

**Table A1**

Ca, Ti, Cr and Mo Isotope anomaly data for various meteorite groups.

| Meteorite Group | $\mu^{48}Ca$ | $\mu^{50}Ti$ | $\mu^{54}Cr$ | $\Delta^{95}Mo$ |
|---|---|---|---|---|
| $NC_{high}$* | 0 ± 5 | 0 ± 2 | 9 ± 13 | 3 ± 15 |
| Enstatite Chondrites | -37 ± 46 | 2 ± 8 | 4 ± 5 | -9 ± 4 |
| Ordinary Chondrites | -28 ± 20 | -66 ± 6 | -40 ± 4 | -15 ± 8 |
| Angrites | -106 ± 33 | -118 ± 8 | -43 ± 6 | -6 ± 9 |
| Ureilites | -182 ± 44 | -185 ± 26 | -92 ± 4 | -15 ± 7 |
| CI Chondrites | 205 ± 20 | 185 ± 12 | 159 ± 6 | 22 ± 34 |
| CAI | 468 ± 205 | 877 ± 53 | 597 ± 52 | 124 ± 14 |

***Note:*** *Data are from Spitzer et al. (2020) for $\mu^{50}Ti$, $\mu^{54}Cr$ and $\Delta^{95}Mo$; Burkhardt et al for $\mu^{48}Ca$ values; Schneider et al. (2020) for $\mu^{54}Cr$ of BSE and $\mu^{48}Ca$ for CAIs are given in Render et al. (2022).*

* *$NC_{high}$ is represented by Bulk Silicate Earth (BSE) for $\mu^{48}Ca$, $\mu^{50}Ti$, $\mu^{54}Cr$ and by NWA 2526 for $\Delta^{95}Mo$ (BSE cannot be used due to mixed composition- Budde et al. 2019)*

**Table A2**

Predicted and measured Sr isotopic compositions.

| Meteorite Group | $f_{IC,NC}$ | Predicted $\mu^{84}Sr$ from (r-,p-)-process variations | Predicted $\mu^{84}Sr$ from s-process variations[1] | Predicted bulk $\mu^{84}Sr$ | Measured $\mu^{84}Sr$ |
|---|---|---|---|---|---|
| Enstatite Chondrites | 0.00 ± 0.01 | -2 ± 4 | 13 ± 4 | 11 ± 6 | 3 ± 11 |
| Ordinary Chondrites | 0.07 ± 0.01 | -8 ± 5 | 21 ± 6 | 13 ± 8 | 4 ± 16 |
| Angrites | 0.11 ± 0.01 | -14 ± 5 | 23 ± 6 | 9 ± 8 | 11 ± 7 |
| Ureilites | 0.16 ± 0.02 | -24 ± 6 | 28 ± 7 | 4 ± 9 | |
| IIAB* | | | 36 ± 9 | | |
| CI | 0.36 ± 0.03 | 32 ± 16 | -12 ± 6 | 20 ± 17 | 20 ± 22 |

* *maximimum NC range*

[1] *Values are calculated by applying a slope of 0.31±0.07 for the s-process $\mu^{84}Sr$-$\mu^{94}Mo$ correlation (Burkhardt et al., 2019) on the range of observed $\mu^{94}Mo$ anomalies. This slope is based on Mo and Sr isotopic data for presolar SiC grains (Nicolussi et al., 1998), acid leachates of primitive chondrites (Burkhardt et al., 2019), and main component s-process yields from Arlandini et al. (1999), using the formalism given by Dauphas et al. (2014). The slope increases to 0.36 if the higher s-process yields from Bisterzo et al. (2014) are used, which would result in a positive shift in the predicted bulk $\mu^{84}Sr$ values of 2, 3, 4, and 4 ppm for the OC, NC, angrites, and ureilites, respectively. This shift is within the uncertainty of our calculations and does not affect the qualitative evidence of our mixing model (Figure 4).*